\documentclass[12pt]{iopart}
\pdfoutput=1

\usepackage[english]{babel}
\expandafter\let\csname equation*\endcsname\relax
\expandafter\let\csname endequation*\endcsname\relax
\usepackage{amsmath}
\usepackage{amssymb}
\usepackage{hyperref}
\hypersetup{urlcolor=blue, colorlinks=true}  
\usepackage{url}

\usepackage{color}
\usepackage{graphicx}
\usepackage[sort&compress,numbers]{natbib}

\newcommand{\be}{\begin{equation}}
\newcommand{\ee}{\end{equation}}

\newcommand{\daga}[1]{#1^{\dagger}}
\newcommand{\Ch}{\mathrm{C} \mathrm{h}}

\newcommand{\ra}{\rangle}
\newcommand{\la}{\langle}
\newcommand{\braket}[1]{\la#1 \ra}

\newcommand{\bibliopath}{HaldaneRef}

\begin{document}

\title{Haldane Model at finite temperature}

\author{Luca Leonforte$^{1,*}$, Davide Valenti$^{1,2}$, Bernardo Spagnolo$^{1,3,4}$, Alexander A. Dubkov$^{3}$ and Angelo Carollo$^{1,3}$}
\address{$^{1}$Department of Physics and Chemistry  ``Emilio Segr\'{e}", Group of Interdisciplinary Theoretical Physics, University of Palermo, Viale delle Scienze, Ed. 18, I-90128 Palermo, Italy}
\address{$^{2}$Istituto di Biomedicina ed Immunologia Molecolare (IBIM) ``Alberto Monroy", CNR, Via Ugo La Malfa 153, I-90146 Palermo, Italy}
\address{$^{3}$Radiophysics Department, National Research Lobachevsky State University of Nizhni Novgorod, 23 Gagarin Avenue, Nizhni Novgorod 603950, Russia}
\address{$^{4}$Istituto Nazionale di Fisica Nucleare, Sezione di Catania, Via S. Sofia 64, I-90123 Catania, Italy} 

\ead{$^*$ luca.leonforte@unipa.it}


\begin{abstract}

We consider the Haldane model, a 2D topological insulator whose phase is defined by the Chern number. 
We study its phases as temperature varies by means of the Uhlmann number, a finite temperature generalization of the Chern number.
Because of the relation between the Uhlmann number and the dynamical transverse conductivity of the system, we evaluate also the conductivity of the model.
This analysis does not show any sign of a phase transition induced by the temperature, nonetheless it gives a better understanding of the fate of the topological phase with the increase of the temperature, and it provides another example of the usefulness of the Uhlmann number as a novel tool to study topological properties at finite temperature.

\end{abstract}



\vspace{2pc}

\submitto{\JSTAT}

\maketitle

\section{Introduction}

The topological ordered phases (TOPs) of matter are a new and interesting field of research that has recently attracted the interest of many physicists. 

Among the TOPs, a subclass of topological phases, the so called symmetric-protected topological phases (SPTPs) were well studied during the last two decades \cite{Ryu2010,Altland1997,Chiu2016,Schnyder2008}. 
These phases are distinguished by the presence or absence of certain discrete symmetries. 
Depending on which symmetries are present, these phases are characterized by different topological invariants.
The presence of a non-trivial topological phase, i.e. a phase that is not topologically equivalent to the vacuum, manifests itself in robust properties of the edge of the system.
Although interesting phenomena have been described from a topological point of view, the systems analyzed are usually considered in their ground state, i.e. at zero temperature. 
The fate of these topological phases with the growing of the temperature is a question that remains unanswered.
To address this point 
various approaches have been proposed to describe the topological phases in mixed state configurations~\cite{Mera2017,Viyuela2014,Budich2015a,He2018}, in the effort of accounting for the effect of temperature at thermal equilibrium or mixedness in
out-of-equilibrium scenarios~\cite{Magazzu2015,Magazzu2016,Spagnolo2018,Valenti2018,Spagnolo2018a,Guarcello2015,Guarcello2016,Spagnolo2016,Spagnolo2015}.
Among these approaches that proposed by Uhlmann to generalize the geometric phase to mixed states seems to be one of the most promising~\cite{Uhlmann1986,Uhlmann1989}.
The Uhlmann geometric phase and its related quantities have been recently used as a starting point to describe the topological phases of systems in a mixed-state configuration \cite{Huang2014,Grusdt2017,Viyuela2014a,Viyuela2014,Mera2017,Linzner2016,Budich2015a,Carollo2018,Carollo2018a,Leonforte2018,Bascone2018}.
Even if 
this tool has lead to successful results in one dimension \cite{Huang2014,Viyuela2014}, the same cannot be said {\color{black} for systems with more dimensions}.
In the case of two dimensions, an interesting approach that follows from the Uhlmann geometric phase has recently given some interesting results \cite{Leonforte2018,He2018}. 
This approach allows to describe the fate of the topological phase with the increase of the temperature, and it also provides a direct link to experimental measurable quantities.
Following this approach in this work, we focus our attention on the well known Haldane model \cite{Haldane1988} which possesses topological non-trivial phases.
This was proposed as a toy model that presents a quantized Hall conductivity, in absence of an external magnetic field, an effect known as anomalous quantum Hall effect (AQHE).
The topological phases for systems topologically equivalent to the Haldane model is captured by the Chern number, that for translational invariant systems is defined as $\Ch = \frac{1}{2 \pi} \iint_{BZ} F_{xy}^B d^2\mathbf{k}$, through the Berry curvature. 
Correspondingly, in the case of thermal equilibrium, through the Uhlmann approach, the \emph{Uhlmann number} $n_U$ has been defined \cite{Leonforte2018}.
$n_U$ is a quantity that describes the evolution of the topological phase as the temperature increases, and that it is linked to the dynamical conductivity of the system.
Focusing therefore on thermal equilibrium, here we study the evolution of both $n_U$ and the dynamical conductivity for the Haldane model.

\section{Topology}

It is well known that a topological order can exist in different classes of systems. 
Among the possible topological orders that a system can have there are the so called symmetry-protected topological phases (SPTPs), which are characterized by the presence of different discrete symmetries (time inversion $\mathcal{T}$, charge conjugation $\mathcal{C}$ and chirality $\mathcal{S}=\mathcal{T} \cdot \mathcal{C}$)\cite{Chiu2016,Ryu2010}. 
The presence or absence of these symmetries allows the classification of matter in ten different classes of systems (ten-fold way) \cite{Ryu2010}.
The class that we're interested in, in this work, is the symmetry class A, which does not posses any of the three discrete symmetries. 
In the case of two-dimensional systems belonging to this class, the topological phases is described by the Chern number, a topological invariant defined 
as the integral over the first Brillouin zone of the Berry curvature:
\begin{equation}
\label{Chern}
\Ch = \frac{1}{2 \pi} \int_{BZ} \mathcal{F}^B_{x y} \rmd k_x \rmd k_y, 
\end{equation}
where the Berry curvature for the ground state is $\mathcal{F}_{\mu \nu}^{B} = \partial_\mu A_\nu^B - \partial_\nu A_\mu^B$, with $A_\mu^B$ the Berry connession defined as $A_\mu^B = i \braket{\psi_0(k) | \partial_\mu | \psi_0(k)}$. 

In the case of two band systems, the Hamiltonian can be put in the following form
\begin{equation}
\label{GenSys}
H = \sum_{\vec{k}} \daga{\Psi}(k) ( \vec{h}(k) \cdot \vec{\sigma} ) \Psi(k),
\end{equation}
where $\Psi(k) = \begin{pmatrix}
a_k  , & b_k 
\end{pmatrix}^T $ is the two component annihilation operator of the two bands. 
In this case the Berry curvature is 
\begin{equation}
\mathcal{F}^B_{x y} = \frac{1}{2} \hat{h(k)} \cdot ( \partial_{k_x} \hat{h} (k) \times \partial_{k_y} \hat{h}(k) ),
\end{equation}
which is the Jacobian of the transformation from the Brilluoin zone to the sphere described by $\hat{h}(k)$. 
The Chern number counts how many times the vector $\hat{h}(k)$ spans the whole sphere moving through the first Brilluoin zone.
This definition of the topological phase it is well defined for system at zero temperature, that is system in its ground state. 
In this case it has been shown that the Chern number is proportional to the transverse conductivity \cite{Laughlin1981,Thouless1983,Thouless1982}
\begin{equation}
\label{TKNN}
\sigma_{x y} = - \frac{e^2}{h} \Ch .
\end{equation}
If the system is in thermal equilibrium, its state will not be anymore the ground state, i.e. a pure state, but it is described by the density operator $\rho = \exp \left( - \beta \hat{H} \right) / \mathcal{Z}$, with $ \mathcal{Z} = \Tr \left[ \exp \left( - \beta \hat{H} \right) \right]$.
As a consequence, in the case of a system in a mixed state, the Berry curvature can not be used to describe the topological phase of the system.
However, thanks to the Uhlmann approach, it is possible to define a geometric phase even for mixed states \cite{Uhlmann1986,Uhlmann1989,Dittmann1999}.

The basis of this approach is to consider an amplitude operator $\omega$ satisfying $\rho= \omega \daga{\omega}$. 
Such a definition leaves a $U(n)$ gauge freedom on the choice of $\omega$, which is the generalization of the $U(1)$ gauge freedom of the pure states, i.e. a phase. 

Let $\rho_\lambda$ be a family of density matrices parametrized by $\lambda \in \mathcal{M}$, with $\gamma := \{ \lambda(t) \in \mathcal{M}, t \in [0,T] \}$ a smooth closed curve in a parameter manifold $\mathcal{M}$ and $\omega_\lambda$ the corresponding path of amplitudes. To reduce the gauge freedom of each amplitude on the curve, Uhlmann introduced a parallel transport condition on $\omega_\lambda$~\cite{Uhlmann1986}.
As in the case of pure states, a cyclic parallel transport over a closed curve $\gamma$ of $\omega_\lambda$ leaves the amplitude unchanged, up to a unitary transformation, $\omega_{\lambda(T)} = \omega_{\lambda(0)} V_\gamma$.
The Uhlmann geometric phase is defined as $\varphi^{U}[\gamma]  = \arg  \Tr
[\daga{\omega_{\lambda(0)}}\omega_{\lambda(T)}]$, which corresponds, in the pure state case, to the phase difference between the initial and final point of a curve along which the pure state is parallel-transported according to the usual Berry connection.

Following the path inaugurated in some recent works \cite{Carollo2018,Leonforte2018}, it is possible to introduce a quantity, the \emph{mean Uhlmann curvature} (MUC), a gauge invariant quantity defined as~\cite{Carollo2018}: 
\begin{equation} 
\mathcal{U}_{\mu \nu} := \lim_{\delta_\mu \delta_\nu \rightarrow 0} \frac{\varphi^{U}[\gamma]}{\delta_\mu
\delta_\nu } = \frac{i}{4} \Tr [ \rho_0 [L_\mu , L_\nu ] ],
\end{equation}
where $L_\mu$ is the symmetric logarithmic derivative, an operator defined implicitly by the following equation
\begin{equation}
\frac{\partial \rho}{\partial \mu} = \frac{1}{2} \{ L_\mu , \rho \}.
\end{equation}
This quantity has the interesting properties that, for a system in a thermal equilibrium, it satisfies:
\begin{equation}
\label{limT}
\lim_{T\rightarrow 0}\mathcal{U}_{\mu \nu} = \mathcal{F}^B_{\mu \nu}.
\end{equation}
In the case of a two-band system in thermal equilibrium, i.e. whose density operator is $\rho=e^{-\beta \mathcal{H}} / \mathcal{Z}$, it can be shown that the MUC satisfies the following expression
\begin{equation}
\mathcal{U}_{\mu \nu} = \tanh \left( \frac{\beta h_k}{2} \right) \tanh^2 \left( \beta h_k \right) \mathcal{F}^B_{\mu \nu},
\end{equation}
where $h_k=|\vec{h}_k|$ is the same quantity used in Eq.(\ref{GenSys}).

The MUC in this form easly verifies Eq.(\ref{limT}).
So, in analogy to the definition of the Chern number, the Uhlmann number is defined for translational invariant systems through the MUC as~\cite{Leonforte2018,Bascone2018}
\begin{equation} 
\label{nu} 
n_U = \frac{1}{2\pi} \int_{BZ} \mathcal{U}_{xy} d k_x d k_y.
\end{equation}
This number, although it is not a topological invariant of the system, tends to the Chern number in the zero temperature limit, and, as it has been shown, it is related to the dynamical conductivity of the system through 
\begin{equation}
\begin{aligned}
\label{nuCond}
n_U \frac{q^2}{2 \pi \hbar} & = -\frac{1}{ \pi } \int_{-\infty}^{+\infty} \frac{d \omega}{\omega} \tanh^2\left( \frac{\hbar \omega \beta}{2} \right) \sigma''_{\mu \nu} (\omega) = \\ 
&= - \frac{1}{2}\int_{-\infty}^{+\infty} d\omega [\sigma^R_{\mu\nu}(\omega)-\sigma^R_{\nu\mu}(\omega)]K_\beta(\omega),
\end{aligned}
\end{equation}
where $ \sigma''_{\mu \nu} (\omega) $ is the transverse dynamical dissipative conductivity and $ \sigma^R_{\mu\nu}(\omega) $ the real part of the transverse conductivity \cite{Leonforte2018}.

The kernel $K_\beta(\omega)$ is a function given by
\begin{align}
K_\beta(\omega)  = \left\{\begin{array}{ll}
\frac{1}{i\pi^{3}}\frac{\Psi^{(1)}\left(\frac{1}{2}-\frac{i \hbar  \beta \omega}{2 \pi }\right)-\Psi^{(1)}\left(\frac{1}{2}+\frac{i \hbar  \beta \omega}{2 \pi }\right)}{\omega}& \omega\neq0 \\[8pt]
-\frac{\hbar \beta}{\pi^{4}} \Psi^{(2)}\left(\frac{1}{2}\right) &\omega=0 \end{array}\right. ,
\end{align}
where $\Psi ^{(n)}(z)$ is the n-th poly-gamma function, defined as $\Psi^{(n)}:=\frac{d^{n+1}}{dz^{n+1}} \ln \Gamma[z]$. 

The kernel $K_\beta(\omega)$ is symmetric, peaked at $\omega=0$, and approximately non-vanishing only within a frequency band $\omega\in\{-\Delta\omega,\Delta\omega \}$ 
of width  $\Delta \omega \simeq \frac{10}{\hbar \beta}$, which provides most of the contributions (about $92\%$) to the integral of Eq.~\eqref{nuCond}.
It can be also proved that the kernel $K_\beta(\omega)$ tends to a delta function in the zero temperature limit
\begin{equation}
\lim_{\beta\to \infty} K_\beta(\omega) =\delta(\omega),
\end{equation}
which shows that Eq.~(\ref{nuCond}) tends to the TKNN formula (see Eq.~(\ref{TKNN})) as temperature vanishes.
\cite{Guarcello2015}
\section{The Haldane Model}
\begin{figure}[!ht]
\center
\includegraphics[scale=2]{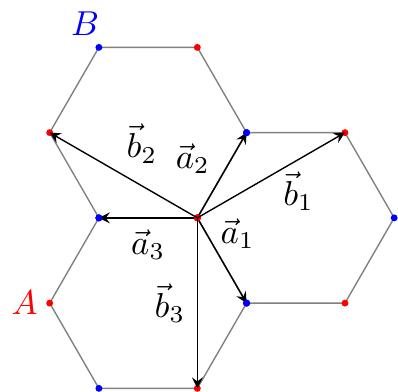}
\caption{Hexagonal cell of the Haldane model, where different colors indicate different sublattices.}
\label{Plaquette}
\end{figure}
The model that we consider here is the well known Haldane model \cite{Haldane1988}, , that represents the prototype of system which displays  the quantum anomalous Hall effect, i.e. a quantized transverse conductivity without an external magnetic field.
The model is made of spinless fermions in a honeycomb lattice with both nearest-neighbor (NN) and next-nearest-neighbor (NNN) interactions. 

The Hamiltonian of this systems is:
\be
H = \sum_i t_0 \daga{c}_i c_i +  \sum_{\langle i,j \rangle } t_1 \daga{c}_i c_j +  \sum_{\langle \langle i,j \rangle \rangle }  t_2 \daga{c}_i c_j .
\ee
As shown in Fig.\ref{Plaquette} the honeycomb lattice can be resolved in two triangular sublattices, that we call $A$ and $B$.
The on-site energy in the two sublattices is $t_{0, A/B} = \pm M$.
The hopping term between nearest-neighbors is equal in all directions and  we will set $t_1=1$.
The hopping term between next-nearest-neighbor is complex and it is defined as $t_2 = t e^{ i \phi }$, where the phase $\phi$ is set, for all terms, to the same value when the hopping occurs clockwise and to the opposite value when the hopping occurs counterclockwise. 
This complex hopping term does not produce a net magnetic flux par unit cell, which means that there is no external magnetic field.
If periodic boundary conditions are considered, then we can write the momentum Hamiltonian as
\begin{equation}
\begin{aligned}
\label{Hk}
H(k) &=  \sum_{j} \left[ 2 t \cos ( \phi ) \cos ( \mathbf{k} \cdot \mathbf{b}_j ) \mathbb{I} + \cos ( \mathbf{k} \cdot \mathbf{a}_j ) \sigma_x + \sin ( \mathbf{k} \cdot \mathbf{a}_j ) \sigma_y \right] +  \\
& + \left\{M - 2 t \sin ( \phi ) \sum_{j} \left[\sin ( \mathbf{k} \cdot \mathbf{b}_j ) \right] \right\}\sigma_z = \\
&= h_0 ( \mathbf{k} ) \mathbb{I} + \vec{h}( \mathbf{k} ) \cdot \vec{\sigma},
\end{aligned}
\end{equation}
where $\mathbf{a}_i$ are defined as the three vectors connecting an A site to a B site, such that $\hat{z}\cdot ( a_i \times a_{i+1})$ is positive, 
with $\mathbf{b}_i=\mathbf{a}_{i+1} -\mathbf{a}_{i-1}$ and $\sigma_i$ the Pauli matrices. 
The Hamiltonian is in the same form as Eq.~(\ref{GenSys}) except for a shift term $h_0(\mathbf{k})$ which is irrelevant to the topology of the system.

In this case the dispersion of the two bands is written as 
\begin{equation}
\epsilon ( \mathbf{k} ) = h_0 ( \mathbf{k} ) \pm |\vec{h}( \mathbf{k} )|,
\end{equation}
where
\begin{equation}
\label{hvec}
|\vec{h}( \mathbf{k} )|^2 = \left( \sum_{j}  \cos ( \mathbf{k} \cdot \mathbf{a}_j ) \right)^2 + \left( \sum_{j} \sin ( \mathbf{k} \cdot \mathbf{a}_j ) \right)^2 + \left( M - 2 t \sin ( \phi ) \sum_{j} \left[\sin ( \mathbf{k} \cdot \mathbf{b}_j ) \right] \right)^2.
\end{equation}
The bands close for those value of $\mathbf{k}$ such that $|\vec{h}( \mathbf{k} )| = 0$, which means that the three terms appearing in Eq.~(\ref{hvec}) must go to zero simultaneously.
This condition is attained only if $M=\pm 3 \sqrt{3} t \sin ( \phi )$.

Finally, by using the bulk Hamiltonian, the Chern number, the topological invariant that distinguish between the different topological phases of the systems, 
can be calculated from Eq. (\ref{Chern}). 
\begin{figure}[!ht]
\center
\includegraphics[scale=1.]{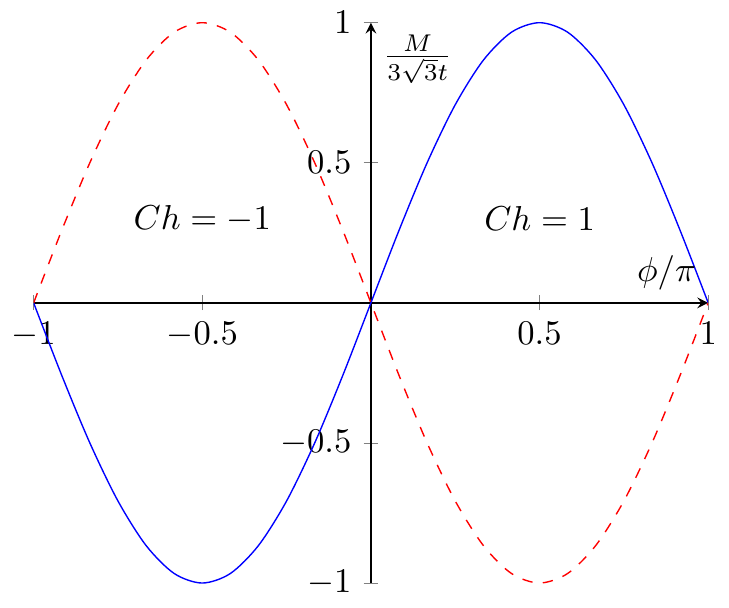}
\caption{Phase diagram of the Haldane model, . Different phases are characterized by different values by the value of the Chern number.}
\label{Phase}
\end{figure}
The phase diagram (see Fig. \ref{Phase}) 
shows that the Chern number is non-zero only if $|M|<3\sqrt{3} t \sin ( \phi )$ and its sign depends on that of $\phi$.
The major difference between this result and the usual integral Quantum Hall effect is that here, in principle, the topological non-trivial effect can be realized without a net external magnetic field.
One may say  therefore that the topological characteristics of the model do not depend on the magnetic field but are intrinsic properties of the band structure of the model itself. 
Due to the complex NNN interactions, many years have passed between the proposition of the model and its experimental realization, which occurred, for the first time, in 2014~\cite{Jotzu2014}. 

\section{Finite Temperature Analysis}

\begin{figure}
\centering
\includegraphics[scale=0.35]{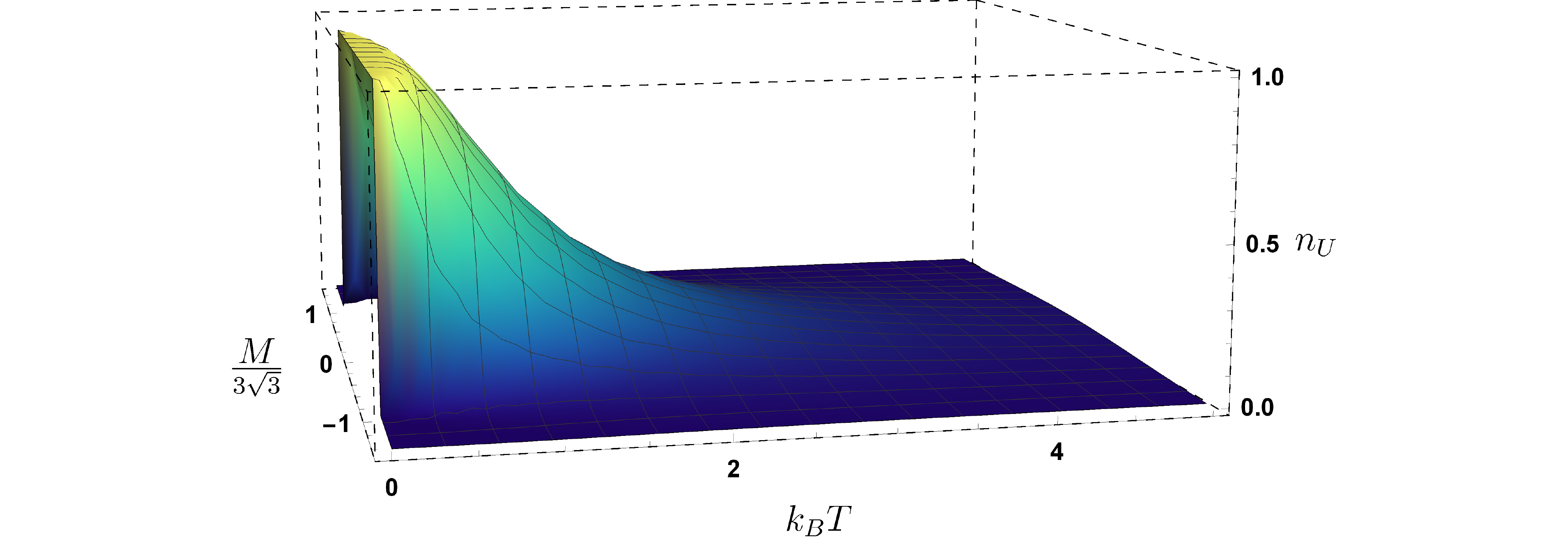}
\caption{Behaviour of the Uhlmann number as a function of $T$ and $M$ with $\phi = \pi /2$.}
\label{nU1}
\end{figure}

\begin{figure}
\label{nU2}
\centering
\includegraphics[scale=0.35]{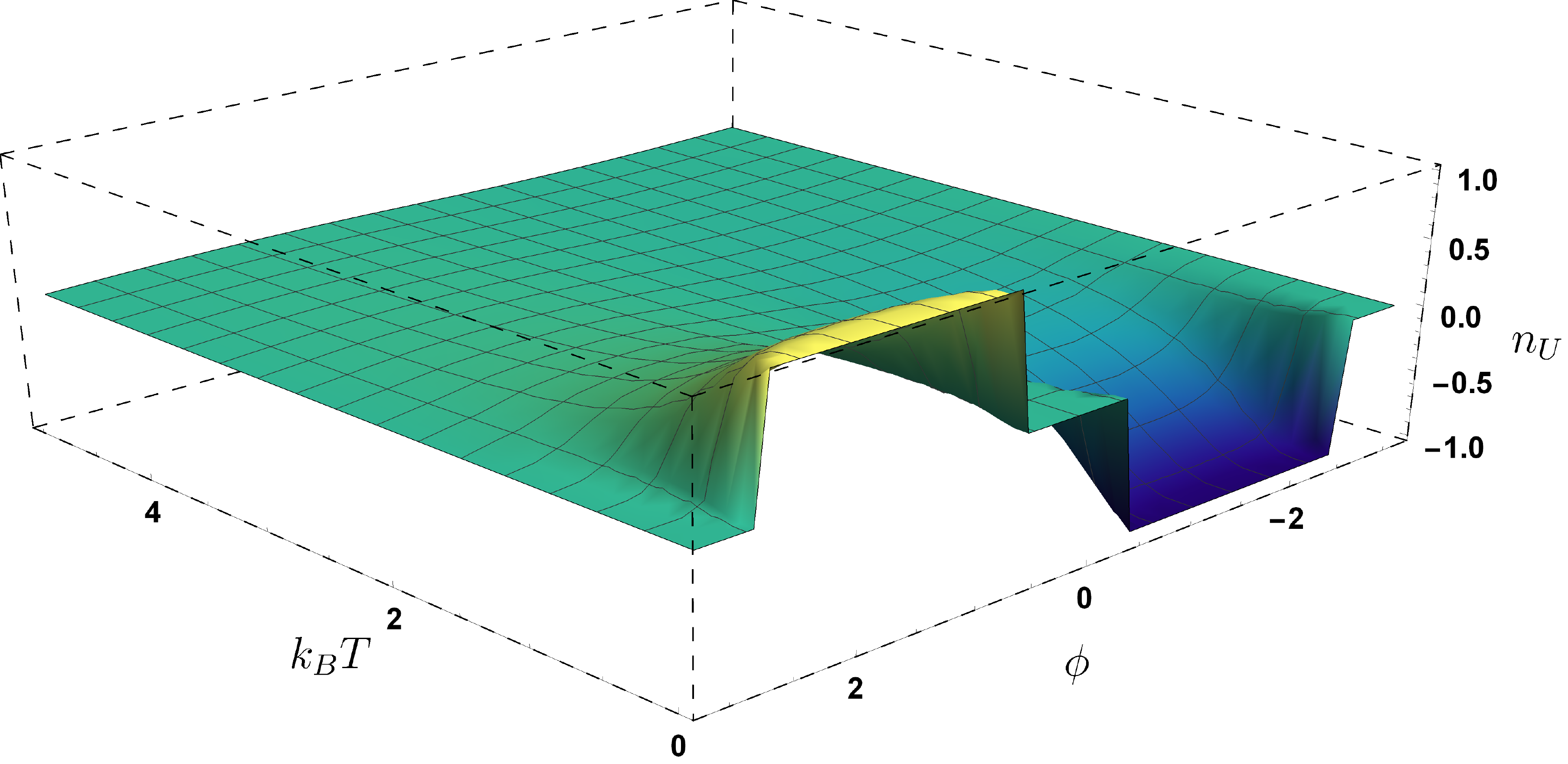}
\caption{ Behaviour of the Uhlmann number as a function of $T$ and $\phi$ with $ M = 3 \sqrt{3} / 2 $.}
\end{figure}

If the system is at thermal equilibrium, the state of the system is not anymore the ground state, but it is described by the density operator $\rho = \exp \left( - \beta \hat{H} \right) / \mathcal{Z}$. 
As previously noted, the description of the topological phase through the Chern number is not valid anymore.
We therefore evaluate the Uhlman number $n_U$, as defined in Eq.~(\ref{nu}).
The results are shown in Fig.~\ref{nU1} and Fig.~\ref{nU2}. 

The behaviour of $n_U$ at zero temperature reproduces the exact results predicted by the Chern number. 
The increase of temperature does not induce any thermal phase transition.
A cross-over behaviour however appears.
Indeed, if we consider, in the parameter space, a point in which the system has a non-trivial topology, i.e. $\Ch \neq 0$, the Uhlmann number vanishes as the temperature increases.
On the other hand in a topological trivial phase, i.e. $\Ch = 0$, the Uhlmann number remains zero, except near the critical point.

\begin{figure}[ht]
\includegraphics[scale=0.58]{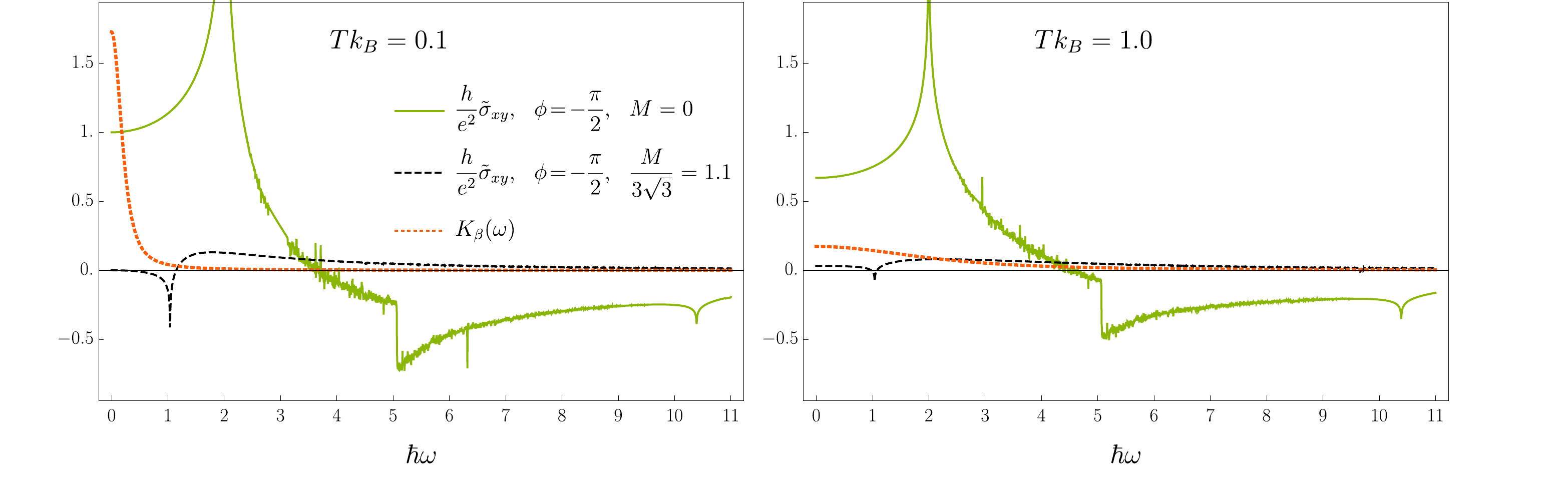}
\caption{ Behaviour of the real transverse conductivity $\tilde{\sigma}_{xy} ( \omega, \beta )$ (in units of $\frac{e^2}{h}$ ) and the Kernel $\mathcal{K}_\beta(\omega)$, 
as a function of the frequency $\omega$ at two different temperatures, $k_B T =0.1$ (left panel) and $k_B T = 1.0$ (right panel), and for two sets of parameters, 
$M=0.$ and $\phi= - \frac{\pi}{2}$ (green line), $\frac{M}{3\sqrt{3}}=1.1$ and $\phi= - \frac{\pi}{2}$ (black line).}
\label{Cond}
\end{figure}

In order to understand better the relation between $n_U$ and the dynamical conductivity (see Eq.~(\ref{nuCond})), in Fig.~\ref{Cond} we show the behaviour of both 
$\tilde{\sigma}_{xy} = \sigma^R_{xy}(\omega)-\sigma^R_{yx}(\omega)$ and the kernel $\mathcal{K}_\beta(\omega)$.

Fig.~\ref{Cond} shows the behaviour of $\tilde{\sigma}_{xy} = \sigma^R_{xy}(\omega)-\sigma^R_{yx}(\omega)$ and the kernel $\mathcal{K}_\beta(\omega)$ as a 
function of the frequency at two different temperatures, $k_B T =0.1$ and $k_B T = 1.0$, and for two sets of parameters.
The green solid line is associated to the set of parameters $M=0$ and $\phi= - \frac{\pi}{2}$, which correspond, at zero temperature, to $\Ch = - 1$. 
The black dashed line is associated to the parameters $\frac{M}{3\sqrt{3}}=1.1$ and $\phi= - \frac{\pi}{2}$, which correspond to a trivial zero temperature Chern number, $\Ch = 0$. 
In particular, we can see that at low temperature the static transverse conductivity approaches the Chern number value, i.e. $ \tilde{\sigma}_{xy} (0) \frac{h}{e^2} = - \Ch$. 
As expected, this behaviour is present also in $n_U$, because of the Dirac delta-like behaviour of  $K_\beta(\omega)$ ( (peaked around $\omega = 0$) at low temperature. 
On the other hand at higher temperature  $K_\beta(\omega)$  broadens out, so that the contribution to $n_U$ comes not only from the static conductivity but also from the 
dynamical conductivity in a broad region of the lower part of the spectrum.

\section{Conclusion}

In this work we studied the topological phase of the Haldane model at finite temperature. 
The phase it is defined at zero temperature through the Chern number. 
To extend this description to the regime of finite temperature we use the Uhlmann number, a quantity directly connected with the Uhlmann geometric phase. 
This quantity differs from its zero temperature counterpart, the Chern number, since it is not a topological invariant. 
Anyway, the Uhlmann number plays a crucial role in the description of the topological properties of a finite temperature system, since it has been shown that it can be linked to the dynamical transverse conductivity of the system.
In this paper we exploited this connection between the Uhlmann number and the transverse conductivity to study the topological properties of the Haldane model. 
While both of these quantities didn't show any temperature induced phase transition, they describe the effect of the temperature on the topological phase and its properties. 
Even in the absence of any phase transition, this analysis shows that non-trivial topological properties are smoothed out by increasing the temperature, and the conductivity at all frequencies is necessary to determine the value of the Uhlmann number. 
This work provides therefore another example of a two-dimensional system, whose topological phase can be described at finite temperature by using the Uhlmann number. 
This analysis can be certainly applied to all the systems which belong to the same class of the Haldane model, and we also expect it can be extended to other two dimensional systems that belong to different topological classes, such as topological superconductors or some time-reversal invariant topological insulators~\cite{Yoshida2012,Yoshida2016,Ivanov2001}.

\section*{Acknowledgments}
This work was supported by the Grant of the Government of the Russian Federation, contract No. 074-02-2018-330 (2). 
We acknowledge also partial support by Ministry of Education, University and Research of the Italian government.

%

\providecommand{\newblock}{}
\bibliographystyle{iopart-num}
\bibliography{\bibliopath}

\end{document}